\documentclass[%
 reprint,
 amsmath,amssymb,
 aps,
]{revtex4-2}

\usepackage{graphicx}
\usepackage{dcolumn}
\usepackage{bm}
\usepackage{color}
\usepackage{amsmath}
\usepackage{wasysym}
\usepackage{float}
\usepackage{array,url,kantlipsum}
\usepackage{verbatim}
\usepackage{xcolor}
\usepackage{CJK}
\usepackage{textcomp}
\newcolumntype{P}[1]{>{\centering\arraybackslash}p{#1}}
\usepackage{array}

\usepackage{hyperref}
\hypersetup{
    colorlinks,
    citecolor=blue,
    filecolor=blue,
    linkcolor=blue,
    urlcolor=blue
}

\newcolumntype{x}[1]{>{\centering\arraybackslash\hspace{0pt}}p{#1}}

\newcommand{\JJ}[1]{\textcolor{orange}{#1}}

\begin{document}



\title{Resonance photoproduction of pionic atoms at the Gamma Factory}

\author{Victor V.\,Flambaum} 
\affiliation{School of Physics, University of New South Wales,  Sydney 2052,  Australia}
\affiliation{Johannes Gutenberg-Universit{\"a}t Mainz, 55128 Mainz, Germany}
 \affiliation{Helmholtz-Institut, GSI Helmholtzzentrum f{\"u}r Schwerionenforschung, 55128 Mainz, Germany}

\author{Junlan Jin}
\affiliation{Department of Modern Physics, University of Science and Technology of China, Hefei 230026, China}

\author{Dmitry Budker}
\affiliation{Johannes Gutenberg-Universit{\"a}t Mainz, 55128 Mainz, Germany}
 \affiliation{Helmholtz-Institut, GSI Helmholtzzentrum f{\"u}r Schwerionenforschung, 55128 Mainz, Germany}
\affiliation{Department of Physics, University of California, Berkeley, California 94720, USA}

\date{\today}
             
\begin{abstract}
We present a possibility of direct resonance production of pionic  atoms (Coulomb bound states of a negative pion  and a nucleus) with a rate  of up to 
$\sim 10^{10}$ 
per second using the gamma-ray beams from the Gamma Factory.

\end{abstract}

\maketitle

\section{Introduction}
The pionic atom \cite{Backenstoss1970_pionic_atom} consists of a negative pion trapped in the Coulomb potential of an atomic nucleus. 
Such systems provide great opportunities to study strong interaction and derive information on nuclear structure. Theoretical study of energy levels in pionic atoms started long ago \cite{Deser1954_En,Brueckner1955_En} and initiated extensive theoretical and experimental studies - see, for example, 
recent experiment on laser spectroscopy of pionic $^4$He of Ref.\,\cite{Hori2020_spectra_helium} and references therein.

A conventional production mechanism of pionic atoms involves creation of free negative pions which are then captured by nuclei. Here, we explore a possibility of direct production of pion-nucleus bound states by a monochromatic gamma-ray beam with the energy tuned to that of the bound state. This monochromatic gamma-ray beam is expected  at the proposed Gamma Factory (GF) \cite{Krasny2015,Budker2020_AdP_GF} currently studied within the CERN Physics Beyond Colliders program. 
It is noted that photoproduction of pionic atoms was once put forward by C. Tzara \cite{TZARA1970_photoproduction_pion}. Here, we introduce different approaches to evaluating photoproduction cross-sections, and 
making use of experimental data for free-pion (both charged and neutral) photoproduction, we extend the analysis to a range of nuclei.
Estimates presented below show that the pionic-atom production rate $p$ may, in principle, reach $\sim 10^{10}$ atoms per second, i.e.  exceed the production rate  at existing facilities 
($\sim 10^{5}$ pionic atoms per second - see, for example, Ref.\,\cite{Hori2020_spectra_helium}) by many orders of magnitude. Specific  experimental arrangements, the discussion of which is beyond the scope of the present  paper,  may 
reduce this gain in the production rate  but the problem certainly deserves a preliminary investigation which is the aim of the present note.

\section{Estimate of the production rate}

We assume an arrangement, where the gamma-rays impinge onto a fixed target. Photoproduction of pionic atoms is realized through the reaction $\gamma+ n \rightarrow p + \pi^-$ within  a nucleus, i.e.,
\begin{equation}
    \gamma + {}^A_{Z_i}\textrm{X} \rightarrow (^A_Z\textrm{X}' +\pi^-)_{nl}\,,
\end{equation}
where ${}^A_{Z_i}$X and $^A_Z\textrm{X}'$ are the initial and final nucleus, respectively (both in their nuclear ground state \footnote{In principle, pionic atoms with final nuclei in excited states can also be resonantly produced with higher photon energies.}), 
$Z_i$ and $Z= Z_i+1$ are the corresponding atomic numbers,  $A$ is the number of nucleons which is the same for the initial nucleus and the final one, $n$ is the pionic atom principal quantum number, and $l$ is the angular momentum quantum number. We will focus mainly on $ns$ states since $\pi^-$ in $ns$ states have larger probability densities inside the nucleus and thus larger cross-sections for their production.

An example of the reaction is
\begin{equation}
\label{HHe}
        \gamma + {}^3\textrm{H} \rightarrow (^3\textrm{He} +\pi^-)_{ns}\,.
\end{equation}
Similar reactions occur with heavier nuclei.
Assuming that the initial atom is at rest, we obtain an equation for the resonant photon energy
$E_{\gamma}= m c^2 + E_{CB} + \Delta \epsilon_{ns} + E_k + \Delta Mc^2$, where  $m c^2=$139.57\,MeV is the rest energy of the negative pion, $c$ is the speed of light,
\begin{equation}
    E_{CB}\approx -\frac{\mu c^2}{2}\left(\frac{Z\alpha}{n}\right)^2
\end{equation}
is the energy of $\pi^-$ in the nuclear  Coulomb field, $\mu=mM/(m+M)$ is the reduced mass, $M$ is the mass of the final nucleus, $\alpha=1/137$ is  the fine structure constant, $\Delta \epsilon_{ns}$ is the energy shift due to the strong interaction (see, for example, Ref.\,\cite{Friedman2007_widths}),
$E_k \approx  p^2/[2(m+M)]$ is the kinetic energy of the pionic atom with momentum  $p=E_{\gamma}/c$, and $\Delta M c^2$ is the change of the mass energy of the final nucleus relative to the initial one. In heavy atoms the relativistic formula for $E_{CB}$ 
should be used \footnote{Note that the relativistic formula for the energy of a spin-zero particle in the case of the point-like nucleus breaks down for $Z\geq 137/2$, so the finite nucleus potential should be used when $Z$ is close to or above  67}. Production of $1s$, $2p$ and several other low-lying $ns$   states in heavy elements, so called deeply bound pionic states, may be especially interesting, since they can provide information on the pion behavior inside the nucleus and cannot be populated via an X-ray cascade in the capture process of a free pion  due to the strong pion absorption in higher states (see, for example, Ref.\,\cite{Friedman2007_widths}). Note that the population of the higher $ns$ states in the cascade is also suppressed since the cascade tends to populate the highest orbital angular momentum (circular) states, $l=n-1$. 
Deeply bound pionic atoms were experimentally studied in a series of measurements such as those in Refs.\,\cite{Yamazaki_1996_deeply_bound_states,Gilg2000_deeply_bound_pionic_207Pb,Geissel2002_deeply_bound_205Pb} via the $^{206,208}$Pb(d,$^3$He) transfer reaction. Other proposals for producing deeply bound pionic atoms 
by methods other than resonance photoproduction are discussed in Refs.\,\cite{Nieves1993_radiative_trapping,Raywood1997_208Pb_radiative_capture,Tain2004_gamma-ray_ALBA}.

Pionic atoms in low-lying atomic states are usually short lived with typical 
lifetimes of $\tau=\hbar/\Gamma_{tot}<10^{-16}$\,s ($\hbar$ is the reduced Planck constant, $\Gamma_{tot}$ is the total width of the pionic state). The negative pion is absorbed by the nucleus, which is followed by
emission of  neutrons, protons, gamma rays, etc.
The produced pionic atoms can
be detected via measuring the fission products following the pion
absorption, where the challenge is to minimize the backgrounds directly induced by the high-energy photons.
We note that detection of gamma rays as a diagnostic for radiative capture of free pions by nuclei was used in Ref.\,\cite{Strandberg2020_threshold_pion_from_deuteron} and, in principle, is also applicable for bound pions in pionic atoms.

\subsection{Photoproduction cross-section for monochromatic photons}
\label{Subsec:monochromatic_cross-section}
The on-resonance cross-section for absorption of monochromatic photons is \cite{Landau1958}
\begin{equation}
\label{Eq:res_cross_section}
    \sigma_0=2 \pi\frac{2 J+1}{2 I+1} \left (\frac{\hbar c}{E_\gamma}\right )^2\frac{\Gamma_{\gamma}}{\Gamma_{tot}},
\end{equation}
where $I$ and $J$ are the spin quantum numbers of the initial and final nucleus, respectively (we assume $I\approx J$ in the following), $E_{\gamma}\approx 140$\,MeV is the photon energy,
$\Gamma_{\gamma}$ is the partial width of the produced pionic state $(^A_Z\textrm{X}' +\pi^-)_{nl}$ corresponding to its 
decay via radiative pion capture with the final nucleus
${}^A_{Z_i}$X in the ground state and $\Gamma_{tot}$ is the total width of the pionic state.
Unfortunately, we are not aware of any calculation of the ratio $\Gamma_{\gamma}/\Gamma_{tot}$. However, it is instructive to start  from a few simple arguments indicating what value we may expect for this ratio. 

The width $\Gamma_{\gamma}$ appears in the Breit-Wigner formula from the entrance channel where the photon is absorbed and the pion is created. The same partial width $\Gamma_{\gamma}$ appears in the decay channel where the pion is absorbed and the photon is emitted. The total width of the pionic atom  $\Gamma_{tot}$ for $ns$ states is dominated by pion absorption (states with large principal quantum number $n$ may be an exception).  The partial width $\Gamma_{\gamma}$ corresponds to an additional photon
radiated in the process of pion absorption. 
The photon emission brings us 
 an extra factor $\alpha$ in the rate of the process. Therefore,  $\Gamma_{\gamma}/\Gamma_{tot} \propto  \alpha = 0.007$ \footnote{Surprisingly, such a naive estimate gives a correct result for the decay $\Delta \to N \pi \gamma$,  where $\Gamma_{\gamma}/\Gamma_{tot}\approx 0.006$ \cite{Zyla2020_Particle_Data} }. 
 The assumption  $\Gamma_{\gamma}/\Gamma_{tot} \sim  \alpha$ would give us $\sigma_0 \sim 1.\times10^{-27}$\,cm$^2$. 
 As we see in Table\,\ref{tab:sigmap_light_nuclei}, this rough estimate is actually valid  for the resonance  cross-section in light nuclei. For heavy nuclei $\sigma_0$ is smaller. 

Indeed, according to Refs.\,\cite{Petrukhin1965_radiative_capture,Delorme1966_radiative_capture}, radiative capture of a pion by the nuclues is typically 2$\%$ of the total capture rate. Here 2$\%$ can only be used as an upper estimate of the ratio $\Gamma_{\gamma}/\Gamma_{tot}$ in Eq.\,\eqref{Eq:res_cross_section}, since after the radiative capture the final nucleus may be not in the ground state (i.e., it could be in an excited or unbound state). For heavier nuclei containing a larger number of nucleons, there are more channels in the radiative capture process so that this ratio may be suppressed significantly.

Though we cannot directly calculate the resonant cross-section for monochromatic photons due to the lack of knowledge of $\Gamma_{\gamma}/\Gamma_{tot}$, we will calculate the integrated cross-section which we actually need for getting production rates of pionic atoms at the GF as discussed below.


\subsection{Integrated cross-section for producing bound pions is related to free-pion production}

For photoproduction of pionic atoms in an $ns$ state with the total width $\Gamma_{tot}(Z,n)$,
the dependence of the absorption cross-section on photon energy is
\begin{equation}
\label{Eq:crosec_energy_dependence_NT}
     \sigma(E)=\sigma_0 \frac{(\Gamma_{tot}/2)^2}{(\Gamma_{tot}/2)^2+(E-E_{\gamma})^2}.
\end{equation}
This gives the integrated cross-section
\begin{equation}
    \int \sigma(E) dE=\frac{\pi}{2} \sigma_0 \Gamma_{tot}(Z,n).
\end{equation}

The cross-section for the production of bound pions may be related to the cross-section $\sigma_p$ for producing free pions with a small pion momentum $p$  and with the final nucleus in its ground state as \cite{TZARA1970_photoproduction_pion}
\begin{equation}
\label{Eq:crosec_bound_pion_v1_NT}
\int \sigma(E) dE=\sigma_p \frac{(Z\alpha)^2}{n^3} K mc^2, 
\end{equation}
where $K= 1-\exp{(-2\pi Z\alpha mc/p)}$. For the  zero pion momentum ($p=0$), the factor becomes $K=1$. 
If we use $\Gamma_{tot}(Z,n)\approx \Gamma_{tot}(Z,1)/n^3$ (see Sec.\,\ref{Subsec:pion_absorption_width}), we have
\begin{equation}
\label{Eq:crosec_bound_pion_NT}
\frac{\pi}{2}\sigma_0 \Gamma_{tot}(Z,1)=\sigma_p (Z\alpha)^2 K mc^2. 
\end{equation}
We see that the resonance cross-section $\sigma_0$ for the $ns$ level production does not depend on $n$, i.e., as expected, $\Gamma_{\gamma}$ and $\Gamma_{tot}$ have the same $n$ dependence - see Eq.\,\eqref{Eq:res_cross_section}.

Note that Eqs.\,\eqref{Eq:crosec_bound_pion_v1_NT} and \eqref{Eq:crosec_bound_pion_NT} are based on the relation between the wavefunctions of a slow free pion and a bound pion at the nucleus. Indeed, in the vicinity of the nucleus the total energy of both the bound and free pions may be neglected in comparison with the large interaction energy. 
Therefore, Schr{\"o}dinger equations for the bound and free pions are the same and their wavefunctions are proportional to each other. The proportionality coefficient is determined from the comparison of the bound and free Coulomb wavefunctions outside the nucleus where there is negligible strong interaction.
This is why the relations of Eqs.\,\eqref{Eq:crosec_bound_pion_v1_NT} and \eqref{Eq:crosec_bound_pion_NT}  are determined by the Coulomb wavefunctions only and are insensitive to the strong interaction. Strictly speaking, this statement is accurate for the wavefunctions with large principal quantum numbers $n$ and small free-pion momenta $p$, which are proportional to the zero-energy solution up to large distances defined by the energy $|E|\ll Ze^2/r$; here $E=E_{CB}$ or $E=p^2/2m$ for bound and free pions, respectively.
However, in atoms with $Z<37$ the results may be extended down to $n=1$ since $1s$ orbital is located mainly outside the nucleus there and the Coulomb-like $1/n^3$ relation between  $ns$ and $1s$ pion densities at the nucleus is reasonably accurate. These relations are inaccurate in heavy atoms with $Z>37$ where $1s$ orbital sits mainly inside the nucleus. Therefore, for the small principal quantum number $n$ and $Z>37$, Eqs.\,\eqref{Eq:crosec_bound_pion_v1_NT} and \eqref{Eq:crosec_bound_pion_NT} may provide only an order-of-magnitude estimate.

\subsection{Cross-section for producing free pions near threshold}
\label{Subsec:free_pion_cross-section}

The  free pion  photoproduction $(\gamma,\pi^-)$ cross-sections have been measured for many nuclei. 
What we need to calculate bound pion production cross-sections is $\sigma_p$ for producing free pions near threshold with the final nucleus directly in its ground state, for which related experimental studies are limited.

Data on $\sigma_{p=0}$ for light nuclei from
Refs.\,\cite{Bernstein1976_pion_threshold_12C_12N,Bosted1979_near_threshold_cross-section,Singham1981_threshold_crosec, Min1976_threshold_11C,Epstein1978_threshold_12N,DeCarlo1980_threshold_14O,Singham1979_threshold_7Li_12C} are listed in Table\,\ref{tab:sigmap_light_nuclei}.

\begin{table}[t]
    \centering
    \begin{tabular}{c c c c c c}
    \hline
    \hline
    ${}_{Z_i}^A$X   &${}_Z^A\textrm{X}'_{g.s.}$   &$\sigma_{p=0}$\,($\mu$b)  &$\sigma_0$\,($\mu$b) &$[10^3\times\Gamma_\gamma/ \Gamma_{tot}]$ &rate (s$^{-1})$\\
    \hline \\[-0.2cm]
        $_3^{7}\textrm{Li}$   &$_4^{7}\textrm{Be}_{g.s.}$   &8     &1200     &9.05     &6.0$\times10^9$\\
        $_5^{11}\textrm{B}$   &$_6^{11}\textrm{C}_{g.s.}$   &4     &260     &1.90     &2.7$\times10^9$\\
        $_6^{12}\textrm{C}$   &$_7^{12}\textrm{N}_{g.s.}$   &4     &200     &1.50     &2.6$\times10^9$\\
        $_7^{14}\textrm{N}$   &$_8^{14}\textrm{O}_{g.s.}$   &0.2   &8   &0.057    &1.3$\times10^8$\\
        
    \hline
    \hline     
    \end{tabular}
    \caption{Parameters of the ${}^A_{Z_i}$X$(\gamma, \pi^-){}^A_Z\textrm{X}'$ reaction for light nuclei. Data for the free $\pi^-$ production at threshold ($\sigma_{p=0}$) in the third column are from \cite{Bernstein1976_pion_threshold_12C_12N,Bosted1979_near_threshold_cross-section,Singham1981_threshold_crosec, Min1976_threshold_11C,Epstein1978_threshold_12N,DeCarlo1980_threshold_14O,Singham1979_threshold_7Li_12C}. Here $^A_Z\textrm{X}'_{g.s.}$ means the final nucleus is in the ground state. $\sigma_0$ is the resonant cross-section for bound $\pi^-$ production. The last column gives the production rate of pionic atom $^A_Z\textrm{X}'$ in $1s$ state expected at the GF by use of Eq.\,\eqref{Eq:max_production_rate_from_free_pion} (see Sec.\,\ref{Subsec:GF_pion_NT}).}
    \label{tab:sigmap_light_nuclei}
\end{table}

Photoproduction of free pions from heavy nuclei was studied by use of a bremsstrahlung-photon beam and radiochemical measurements, for example, in Refs.\,\cite{Sakamoto1989_51Cr_from_51V,Sakamoto1990_133Ba_from_133Cs,Blomqvist1978_197Hg_from_197Au,Sakamoto1999_complex_nuclei}.
According to Ref.\,\cite{Sakamoto1999_complex_nuclei}, yields for $(\gamma, \pi^-)$ from nuclei with $A > 40$ are almost $A$-independent at bremsstrahlung end-point energies $E_0$=250\,MeV, 400\,MeV and 800\,MeV. In  Ref.\,\cite{Sakamoto1999_complex_nuclei}, the authors did not use quasi-monoenergetic gamma-rays with near-threshold energy and the final nuclei could be in excited states instead of only the ground state, so we can only take $\sigma_{p=0}$ to be $Z$-independent for nuclei with $Z\geq 20$ as a rough estimate.
Calculations of $\sigma_p$ for the $^{197}$Au($\gamma,\pi^-$)$^{197}$Hg$_{g.s.}$ reaction in Ref.\,\cite{Blomqvist1978_197Hg_from_197Au} produce $\sigma_p\approx 1\,\mu$b at $p\approx92\,$MeV$/c$ while experimental data presented  in Ref.\,\cite{Blomqvist1978_197Hg_from_197Au} show a significantly larger result, up to  $\sigma_p\sim 100\,\mu$b. However, radiochemical measurements are slow, therefore  all excited nuclei have enough time to decay to the ground state or to metastable isomeric state, so only these two outcomes, accumulated ground state or accumulated isomeric state, have been experimentally separated.  
Therefore, we suggest $\sigma_{p=0}\approx 1\,\mu$b as an order-of-magnitude estimate for producing free pions from  $^{197}$Au near threshold with produced $^{197}$Hg directly in its ground state. Together with the $Z$-independence noted above, we apply the estimate  $\sigma_{p=0}(Z)\approx 1\,\mu$b 
to $9\leq Z\leq 92$ cases.

Cross-sections for $(\gamma, \pi^-)$ reaction can also be estimated using experimental data on $(\gamma, \pi^0)$ reaction.
An advantage of  $(\gamma, \pi^0)$ reaction is  that in photoproduction of $\pi^0$ with the final nucleus in the ground state same as the initial nucleus, there is coherent enhancement $\sim A$ in  the production cross-sections. This means that in the near-threshold area we do not need to worry about other channels with excited nuclei in the final states, which are not enhanced. Coherent  enhancement is strongly suppressed in photoproduction of $\pi^-$ with a neutron in the initial nucleus converted into a proton. This gives a factor $(A-Z)/A^2\approx 1/(4Z)$ in the ratio of  $\sigma_p(\pi^-)/\sigma_p(\pi^0)$ \footnote{There are indications that, due to pion repulsion from the nucleus only surface nucleons contribute to the coherent cross-section, see, for example, Ref.\,\cite{Boffi1991_coherent_pion0}. If this is the case, we may estimate the suppression factor as $1/(4Z)^{2/3}$.}.    
Coulomb interaction between the nucleus and $\pi^-$ adds the Sommerfeld factor \cite{TZARA1970_photoproduction_pion} to the cross-section near the threshold. Therefore, for producing $\pi^-$ and $\pi^0$ at the same momentum we get 
\begin{equation}
\label{Eq:neutral_pion}
    \sigma_p(\pi^-)\approx \frac{1}{4Z} \frac{2\pi f}{1-\exp{(-2\pi f)}}\sigma_p(\pi^0),
\end{equation}
where $f= Z\alpha mc/p$ and $m$ is the mass of $\pi^-$.
Note that in this relation momenta of $\pi^-$ and  $\pi^0$ are the same but photon energies for producing them are different since masses of $\pi^-$ and  $\pi^0$ are different. 
With cross-section data for coherent photoproduction of $\pi^0$ read from figures in Ref.\,\cite{Boffi1991_coherent_pion0} (at $E_\gamma\approx 146$\,MeV), we get for producing $\pi^-$ at threshold [$\sigma_{p=0}(\pi^-)= \sigma_p(\pi^-) K$, see Eq.\,(\ref{Eq:crosec_bound_pion_v1_NT})] $\sigma_{p=0}(Z=6)\approx 1.0$\,$\mu$b,
$\sigma_{p=0}(Z=20)\approx 3.2$\,$\mu$b, and $\sigma_{p=0}(Z=82)\approx 3.5$\,$\mu$b.
Within expected order-of-magnitude accuracy these estimates are comparable to the estimate presented above, $\sigma_{p=0}(Z)\approx 1\,\mu$b for $9\leq Z\leq 92$.

\begin{figure}[!htpb]\centering
    \includegraphics[width=1.0\linewidth]{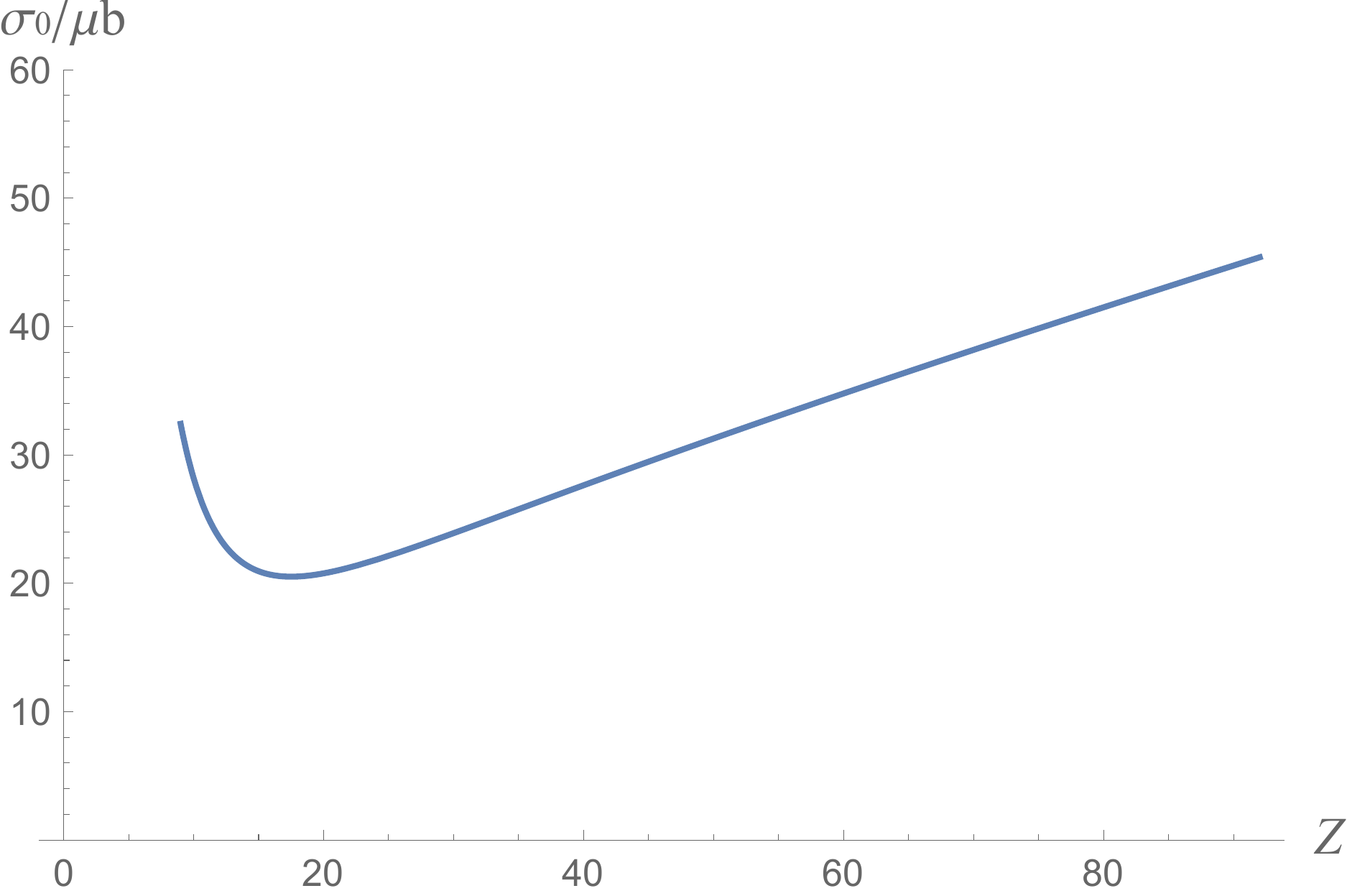}
    \caption{Resonant cross-section $\sigma_0$ for producing $ns$ pionic states with final nuclei with $9\leq Z\leq 92$.}
    \label{fig:sigma0}
\end{figure}

\begin{figure}[!htpb]\centering
    \includegraphics[width=1.0\linewidth]{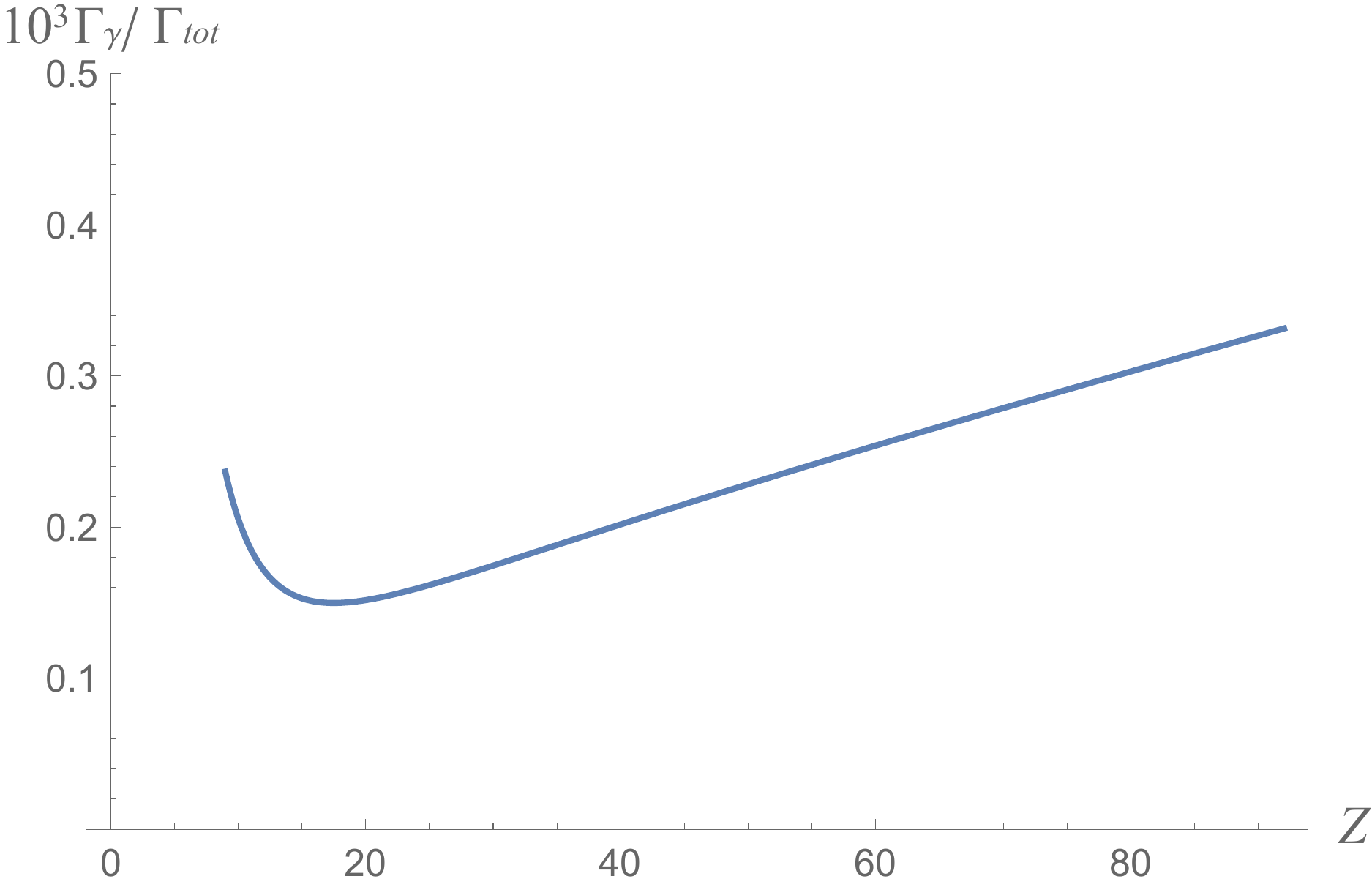}
    \caption{Width ratio $\Gamma_\gamma/ \Gamma_{tot}$ (multiplied by $10^3$) of $ns$ pionic states with final nuclei with $9\leq Z\leq 92$.}
    \label{fig:branching_ratio}
\end{figure}

\begin{figure}[!htpb]\centering
    \includegraphics[width=1.0\linewidth]{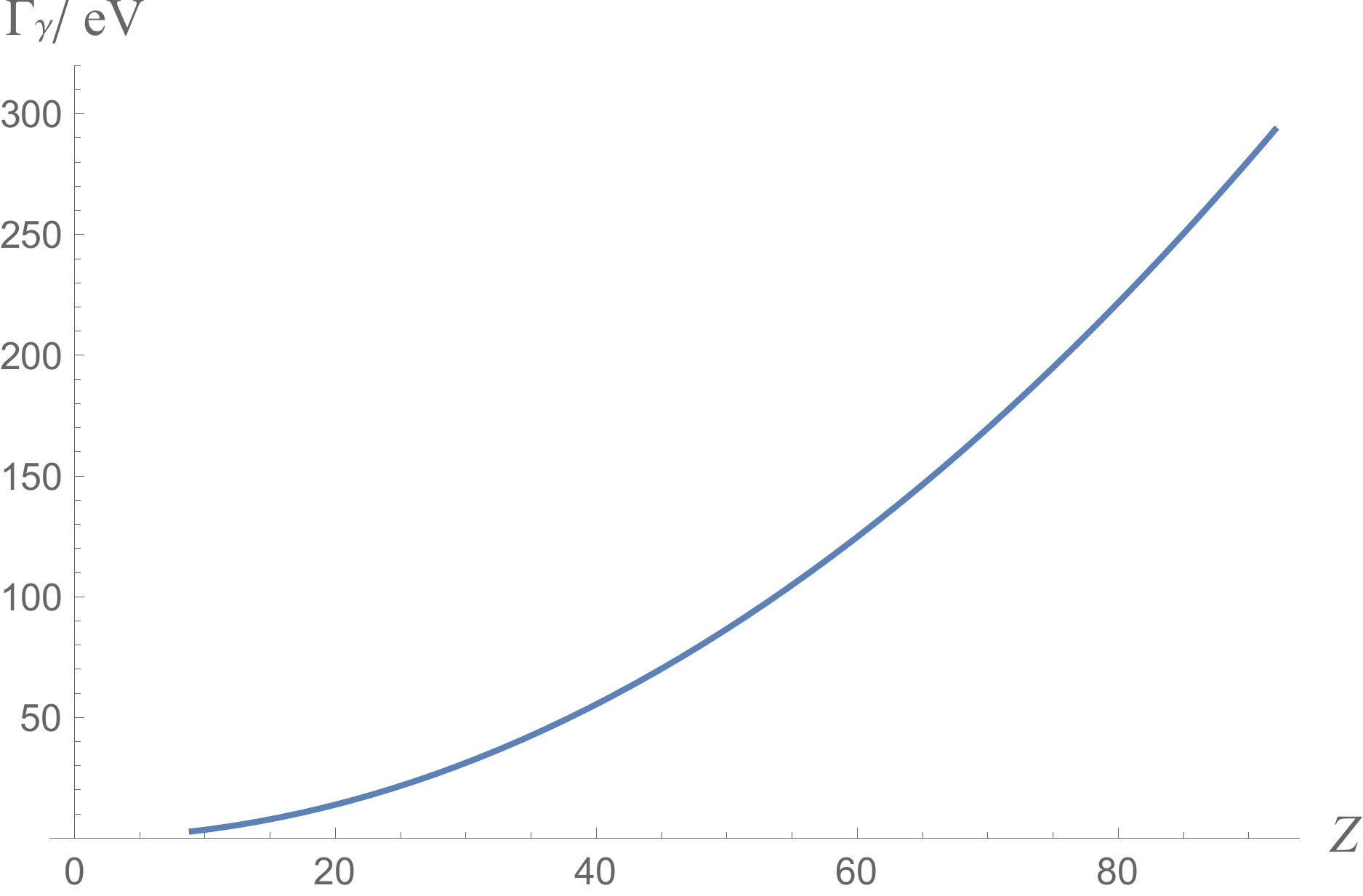}
    \caption{Partial width $\Gamma_\gamma$ of the $1s$ pionic state with final nuclei with $9\leq Z\leq 92$.}
    \label{fig:partial_width}
\end{figure}

Using $\sigma_{p=0}\approx 1\,\mu$b (for $9\leq Z\leq 92$) and $\Gamma_{tot}(Z,1)$ (Sec.\,\ref{Subsec:pion_absorption_width}), we can get the resonant cross-section $\sigma_0$, the width ratio $\Gamma_\gamma/ \Gamma_{tot}$ and the partial width $\Gamma_\gamma$ as shown in Figs.\,\ref{fig:sigma0}, \ref{fig:branching_ratio} and \ref{fig:partial_width}, respectively. 
Figs.\,\ref{fig:sigma0} and \ref{fig:branching_ratio}
look similar since $\sigma_0$ is proportional to $\Gamma_\gamma/ \Gamma_{tot}$ according to Eq.\,\eqref{Eq:res_cross_section}. For light nuclei with $Z\leq 8$, $\sigma_0$ and $\Gamma_\gamma/ \Gamma_{tot}$ are obtained with corresponding $\sigma_{p=0}$ data  - see Table\,\ref{tab:sigmap_light_nuclei}.

With known $\sigma_{p=0}$, integrated cross-sections for producing pionic atoms in $ns$ states can be derived directly using Eq.\,\eqref{Eq:crosec_bound_pion_v1_NT}. Then we can discuss production rates expected at the GF \cite{Krasny2015,Budker2020_AdP_GF}.

\subsection{Production rate with GF photons}
\label{Subsec:GF_pion_NT}
At the GF, laser photons will be  sent
onto an incoming beam of partially stripped ions (PSI) with a high relativistic factor $\gamma$. Here $\gamma= 1/\sqrt{1-\beta^2}$ can be set between $\gamma\approx 200$ and $\gamma\approx 3000$ (in the case of the LHC ring), where $\beta=v/c\approx 1$ and $v$ is the speed of the ions. In the ion frame, the energy of incident photons is boosted by a factor of $\approx2\gamma$ and atomic-level transitions in the PSI can be excited. The secondary photons emitted by the excited ions, are seen in the lab frame to be predominantly emitted in the direction of propagation of the PSI. Their energy is boosted by another factor of up to $\approx 2\gamma$, and can be tuned by changing $\gamma$ and the energy of the laser photons.

Neglecting the energy spread of the ions which is $\approx 10^{-4}$ and can be reduced further to $\approx 10^{-6}$ as projected with the application of laser cooling of the PSI, secondary photons at the GF are almost uniformly distributed up to the maximal energy. It will be tuned to $E_{\gamma}\approx 140$\,MeV for producing pionic atoms. Because the energy of 
those photons is correlated with their propagation angle, after proper collimation, the effective photon flux can be estimated as 
\begin{equation}
    j_{eff}\approx j \frac{\Gamma_{ph}}{E_{\gamma}},
\end{equation}
where $j= 10^{17}$\,s$^{-1}$ is the expected total photon flux at the GF and $\Gamma_{ph}$ is the photon-distribution width of GF photons after the collimation. We can see that a reduction in $\Gamma_{ph}$ comes at a cost of the effective photon flux.
After the collimation, GF photons are nearly uniformly distributed between $E_{\gamma}-\Gamma_{ph}/2$ and $E_{\gamma}+\Gamma_{ph}/2$.
Considering the photon-distribution width, the effective photon-absorption cross-section of the initial nucleus for creating pionic atoms in the $ns$ state with the atomic number $Z$ is
\begin{equation}
\label{cutoff_NT}
    \sigma_{eff}(Z_i,n)= 
    \sigma_0 \frac{\Gamma_{tot}(Z,n)}{\Gamma_{ph}}\arctan{\frac{\Gamma_{ph}}{\Gamma_{tot}(Z,n)}}.
\end{equation}
To make better use of photon fluxes at the GF, i.e., get larger $j_{eff}\sigma_{eff}$, we should choose $\Gamma_{ph}\geq \Gamma_{tot}(Z,n)$, which leads to $\sigma_{eff}(Z_i,n)\approx \sigma_0\Gamma_{tot}(Z,n)/\Gamma_{ph}$.

When producing pionic atoms with 140\,MeV photons, there are other competing processes. Various contributions to the photon attenuation cross-section are shown in Ref.\,\cite{Zyla2020_Particle_Data}. In our case,  pair production in the nuclear field and Compton scattering off electrons are dominant.
According to Ref.\,\cite{Berestetskii_QED}, 
the cross-section due to scattering of 140\,MeV photons by electrons is 
\begin{equation}
    \sigma_{scat} \approx 2\pi Z_i r_e^2 \frac{m_e c^2}{2E_{\gamma}}\left(\ln{\frac{2E_{\gamma}}{m_e c^2}}+\frac{1}{2}\right),
\end{equation}
and the cross-section due to pair production in the nuclear field is
\begin{equation}
\label{Eq:nuclear_pair_production_NT}
    \sigma_{pp} \approx \frac{28}{9} Z_i^2 \alpha r_e^2 \left(\ln{\frac{2E_{\gamma}}{m_e c^2}}-\frac{109}{42}\right),
\end{equation}
where 
$r_e = 2.8$\,fm is the classical electron radius, and $m_ec^2=0.511$\,MeV is the rest energy of an electron.
Thus, we estimate the total cross-section as
\begin{equation}
    \sigma_{tot}(Z_i)\approx \sigma_{scat}+\sigma_{pp}\approx 0.6\times10^{-26}(Z_i^2+Z_i)\,\textrm{cm}^2.
\end{equation}

When all GF photons (after collimation) are absorbed in the target, the maximal production rate is 
\begin{equation}
\label{Eq:max_production_rate_NT}
 p_{max}(Z,n) \approx j_{eff} \frac{\sigma_{eff}(Z_i,n)}{\sigma_{tot}(Z_i)}= j \frac{\Gamma_{tot}(Z,n)}{E_{\gamma}}\frac{\sigma_0}{\sigma_{tot}(Z_i)}.
\end{equation}
Substituting $\sigma_0$ from Eq.\,\eqref{Eq:res_cross_section} or the integrated cross-section from Eq.\,\eqref{Eq:crosec_bound_pion_NT}, we get 
\begin{equation}
\label{Eq:max_production_rate_from_partial_width}
p_{max}(Z,n) \approx 
2 \pi\frac{2 J+1}{2 I+1}\frac{\Gamma_{\gamma}(Z,n)}{E_{\gamma}}
\frac{(\hbar c/E_{\gamma})^2}{\sigma_{tot}(Z_i)}j
\end{equation}
and 
\begin{equation}
\label{Eq:max_production_rate_from_free_pion}
p_{max}(Z,n)\approx \frac{2}{\pi} \frac{\sigma_{p=0}(Z)}{\sigma_{tot}(Z_i)} \frac{Z^2 \alpha^2}{n^3}j\approx 5.6\times10^8 \frac{\sigma_{p=0}(Z)/1\mu\textrm{b}}{n^3}\,\textrm{s}^{-1}
\end{equation}
\footnote{There is an extra coefficient $Z/(Z-1)$ in Eq.\,\eqref{Eq:max_production_rate_from_free_pion} that should be taken into account for light nuclei.}.
The production rate obtained with Eq.\,\eqref{Eq:max_production_rate_from_free_pion} is proportional to $\sigma_{p=0}$. 
With the approximation 
$\sigma_{p=0}(Z)\approx 1\,\mu$b for $9\leq Z\leq 92$, we get
$p_{max}(Z,1)\approx 5.6\times10^8$\,s$^{-1}$. Production rates for light nuclei in $1s$ states are presented in  Table\,\ref{tab:sigmap_light_nuclei}.


The production rate could reach up to $\sim 10^{10}$ per second. 
Such high production rate of pionic atoms would be a significant improvement compared with other production methods (see, for example, Ref.\,\cite{Hori2020_spectra_helium} where $\sim 10^5$ pionic helium atoms per second are produced at a 590\,MeV proton facility).

To suppress the background induced by high-energy photons, $\Gamma_{ph}$ should be reduced by collimation but remain larger than $\Gamma_{tot}$ to obtain high production rates (the total widths for $ns$ pionic states $\Gamma_{tot}$ are discussed in Sec.\,\ref{Subsec:pion_absorption_width}). Therefore, $\Gamma_{ph}=\Gamma_{tot}$ is a good choice of $\Gamma_{ph}$ except when $\Gamma_{tot}$ is smaller than the lower limit of $\Gamma_{ph}$ which could be $\approx$140\,eV at the GF. It is also possible to produce pionic atoms in various bound states at the same time utilizing more photons and achieving even higher total production rates.
To realize this we can tune 
$\Gamma_{ph}\approx E_{1s}(Z)$, where $E_{1s}$ is the $1s$ binding energy of the pionic atom (see, for example, Refs.\,\cite{Toki1989_pion_width_1s,Friedman2007_widths}).

\subsection{Estimate of the pion absorption width} 
\label{Subsec:pion_absorption_width}

\begin{figure}[!htpb]\centering
    \includegraphics[width=1.0\linewidth]{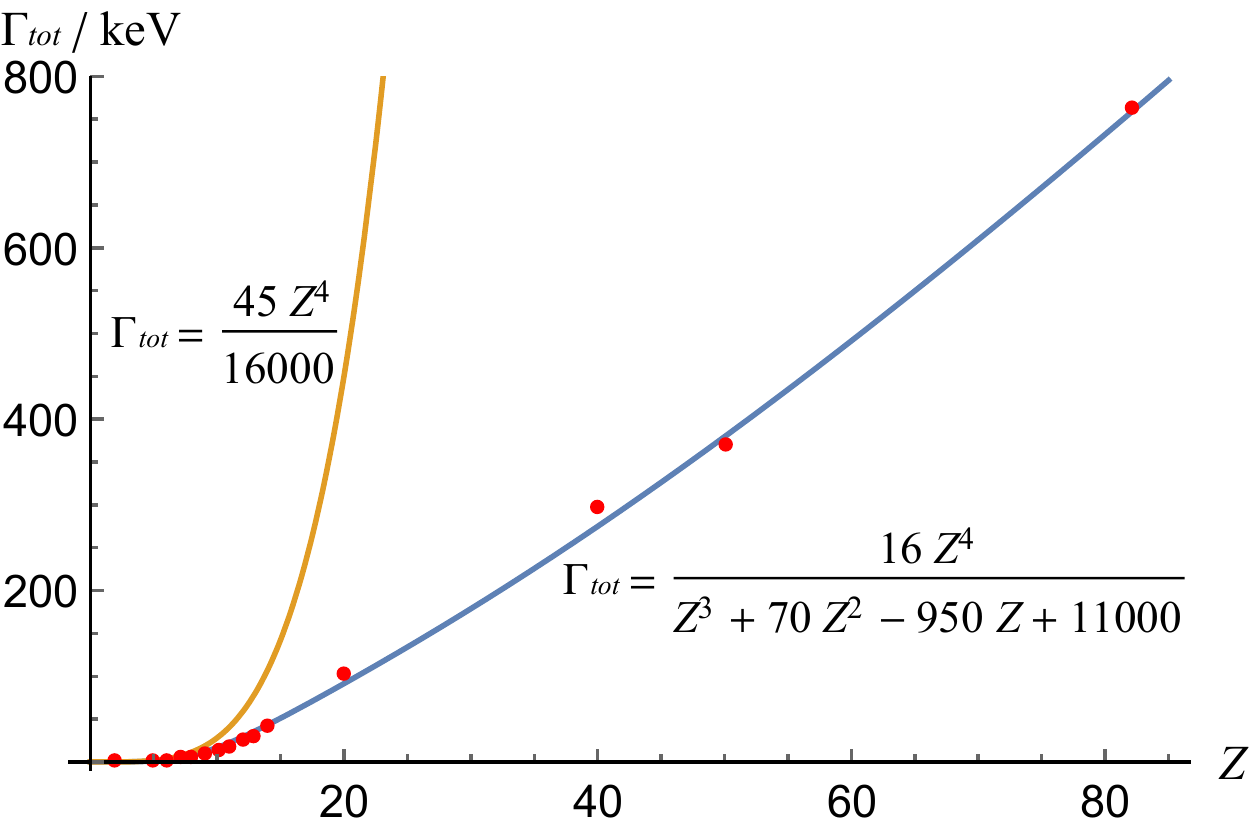}
    \caption{Fitting of $1s$ width data from Refs.\,\cite{Konijn1990_widths_nl,Friedman2007_widths} shown as red dots on the graph. When width data for different isotopes with the same atomic number $Z$ are given, the average value of the widths is used for the fitting. Data for $^{40}$Ca and $^{90}$Zr are from Ref.\,\cite{Toki1989_pion_width_1s}. $1s$ width for $^4$He is from Ref.\,\cite{Backenstoss1974_X-ray_helium}. The yellow curve and blue curve represent Eq.\,\eqref{Eq:Width_approx} and Eq.\,\eqref{Eq:1s_width}, respectively.}
    \label{fig:width_fitting}
\end{figure}

The width of the atomic energy level $\Gamma_{tot}$ for $ns$ states is dominated by pion absorption.
As an approximation we may assume that $\Gamma_{tot}$ is proportional to $Z^4/n^3$, where $Z^3/n^3$ is from the probability density of the $ns$-wavefunction at the origin , $\psi_{ns}^2(0)$, and the extra factor $Z$ appears if each proton may absorb the pion.
According to Ref.\,\cite{Backenstoss1974_X-ray_helium}, $\Gamma_{tot}$ of $(^4\textrm{He} +\pi^-)_{1s}$ is 45\,eV. This gives us an estimate of the absorption width in the vicinity of $Z= 2$: 
\begin{equation}
\label{Eq:Width_approx}
    \Gamma_{tot}(Z,n)\approx 45\frac{Z^4}{16n^3}\,\textrm{eV}.
\end{equation}
However, this formula strongly overestimates pion absorption width for $Z \gg 1$. Due to the repulsive strong interaction between the pion in $1s$ state and the nucleus,
the increase of $\Gamma_{tot}$ could be small as $Z$ increases \cite{Friedman_1985_width_heavy_atoms}. Thus the formula above needs to be adjusted. 
By fitting  data for the $1s$ level width of pionic atoms with $Z$ up to 82 from Refs.\,\cite{Toki1989_pion_width_1s,Konijn1990_widths_nl,Friedman2007_widths} (see Fig.\,\ref{fig:width_fitting}), we approximate the pion absorption width of the $1s$ level as

\begin{equation}
\label{Eq:1s_width}
    \Gamma_{tot}(Z,n=1)\approx \frac{16Z^4}{Z^3+70Z^2-950Z+11000}\,\textrm{keV}.
\end{equation}

For $ns$ states with $n\geq2$, which have smaller widths and longer lifetimes, 
the pion absorption width can be estimated as $\Gamma_{ns}=\Gamma_{1s}/n^3$ 
The $1/n^3$ dependence for the width is violated in heavy atoms since the pion in the $1s$ state of heavy atoms has orbital radius close to the nuclear radius and the nucleus cannot be approximated as a point-like center. 
For pionic atoms in $ns$ states with $Z/n^{3/2} \leq 37$ the radius of the corresponding Bohr orbit [$\approx 200 n^2/Z$ fm]  is larger than the nucleus [$\approx1.2(2.5Z)^{1/3}$\,fm]. For $Z\leq 37$ 
we estimate the width $\Gamma_{tot}(Z,n)$ as
\begin{equation}
\label{Eq:ns_width}
    \Gamma_{tot}(Z,n)\approx \Gamma_{tot}(Z,1)\frac{1}{n^3}.
\end{equation}
For pionic atoms with  $Z >37$ this estimate of the total width $\Gamma_{ns}$ is not applicable since the $1/n^3$ dependence follows from the Coulomb wavefunctions but $1s$ function in this case is not the Coulomb one. Therefore,  for $Z >37$ we may use Eq.\,\eqref{Eq:ns_width} only as an  order-of-magnitude estimate for  $\Gamma_{tot}(Z,n)$.

Besides $ns$ states, pionic atoms in other $nl$ states can also be produced.
When the radii of corresponding Bohr orbits exceed the nuclear size,
the widths of these states can be estimated as $\Gamma_{nl}/\Gamma_{ns}\approx R_{nl}(r_0)^2/ R_{ns}(r_0)^2$. Here $R_{nl}$ is the radial wavefunction of hydrogen-like atoms, and  $R_{nl}(r_0)^2/ R_{ns}(r_0)^2$ gives the ratio of probabilities of a pion in $nl$ and $ns$ state to occur on the nuclear surface. For an estimate, $\Gamma_{nl}/\Gamma_{ns}\sim (Z r_0/a_0)^{2l}$, where $r_0\approx 1.2 (2.5Z)^{1/3}$\,fm is the nuclear radius, and $a_0\approx 200$\,fm is the $1s$ Bohr radius of the pionic hydrogen atom. There are some $nl$-level width data for pionic atoms in Ref.\,\cite{Friedman2007_widths}, and we see that the simple  estimate $\Gamma_{nl}/\Gamma_{ns}\sim (Z r_0/a_0)^{2l}$ is consistent with the data for $2p$-level widths within one order of magnitude.
The widths for $3d$ and $4f$ states should be estimated using more accurate  expression $\Gamma_{nl}/\Gamma_{ns}\approx R_{nl}(r_0)^2/ R_{ns}(r_0)^2$.

Atomic levels also have a radiative width. Following the calculation for hydrogen \cite{BetheSalpeter} we obtain the following estimate of the radiative width for the pionic $ns$-levels with $n>2$: $\Gamma_{rad} \approx 10^{-4} Z^4/n^3$\,eV. This width is much smaller than the pion-absorption width. Note that for large $n$ the width due to emission of Auger electrons may exceed the radiative width. However, for $n <\sqrt{m_\pi/ m_e}\approx 16$ the pionic atom de-excites primarily by an X-ray cascade, which has electric dipole character to a good approximation - see, for example, Ref.\,\cite{Ericson1988}.

\section{Conclusion}

Estimates presented above indicate that the direct resonance photoproduction of pionic atoms in $ns$ states at the Gamma Factory would provide several orders of magnitude larger number of these atoms than any existing facility. The $ns$ states are especially sensitive and important for the study of the interaction between the pion and nucleons, nuclear structure and nuclear forces forming the structure, including, for example, the neutron skin problem related to the prediction of the neutron-star equation of state and maximal neutron-star mass.
This study of the strong interaction effects is, in fact, the main aim of 
the pion-atom production. Production of low-lying states in heavy elements, so called deeply bound pionic states, may be especially interesting, since they can provide information on the pion behavior inside the nucleus and cannot be populated via an X-ray cascade in the capture process of a free pion   due to the strong pion absorption in higher states. Note that the population of the higher $ns$ states via an X-ray cascade is also suppressed since the cascade tends to 
populate the highest orbital angular momentum (circular) states, $l=n-1$ (see, for example, Ref.\,\cite{Friedman2007_widths}).  


\section*{Acknowledgments}
The authors are grateful to Carlos Bertulani, Catalina Curceanu, Eli Friedman, Avraham Gal, Mikhail Gorshteyn, Masaki Hori, M. Witold Krasny, and Vladimir G.\,Zelevinsky  for helpful information and discussions. The work of  VF is supported by the Australian Research Council grants DP190100974 and DP20010015 and the Gutenberg Fellowship. The work of DB was  supported in part by the DFG Project ID 390831469:  EXC 2118 (PRISMA+ Cluster of Excellence).

\bibliography{Pionic_Atoms}

\end{document}